%% file: mainfile.tex
\documentstyle[12pt,aps,pre,psfig]{revtex}

\newcommand {\be}{\begin{equation}}
\newcommand {\ee}{\end{equation}}
\newcommand {\bea}{\begin{eqnarray}}
\newcommand {\eea}{\end{eqnarray}}

\begin{document}
\title{Identification and characterization of systems with delayed
feedback: (I) Theory and tools}
\author{Martin J.\ B\"unner$^1$, Marco Ciofini$^1$, Antonino Giaquinta$^2$, 
Rainer Hegger$^3$}
\author{Holger Kantz$^3$, Riccardo Meucci$^1$ and Antonio Politi$^1$}
\address{(1) Istituto Nazionale di Ottica\\ Largo E.\ Fermi 6,
50125 Firenze, Italy}
\address{(2) Instituto Pluridisciplinar, Paseo Juan XXIII 1 -- 28040
Madrid, Spain}
\address{(3) Max--Planck--Institut f\"ur Physik komplexer Systeme\\
N\"othnitzer Str.\ 38, 01187 Dresden, Germany}

\date{\today}
\maketitle

\begin{abstract}
High-dimensional chaos displayed by multi-component systems with a
single time-delayed feedback is shown to be accessible to time series 
analysis of a scalar variable only. The mapping of the original dynamics
onto scalar time-delay systems defined on sufficiently high dimensional
spaces is thoroughly discussed. The dimension of the ``embedding'' space
turns out to be independent of the delay time and thus of the dimensionality
of the attractor dynamics. As a consequence, the procedure described in
the present paper turns out to be definitely advantageous with respect to 
the standard ``embedding'' technique in the case of high-dimensional chaos, 
when the latter is practically unapplicable. The mapping is not exact when 
delayed maps are used to reproduce the dynamics of time-continuous systems,
but the errors can be kept under control. In this context, the approximation 
of delay-differential equations is discussed with reference to different classes 
of maps. Appropriate tools to estimate the a priori unknown delay time and 
the number of hidden components are introduced. The generalized Mackey-Glass 
system is investigated in detail as a testing ground for the theoretical 
considerations.
\end{abstract}

\pacs{\hspace{1.9cm}  PACS numbers: 02.30.Ks,05.45.Tp}

\section{Introduction}
\input{introduction.tex}

\section{Embedding theory}
\input{model1.tex}

\section{From continuous to discrete time}
\input{model2.tex}

\section{Local indicators}
\input{implementation.tex}

\section{Global indicators}

\input{m-estimation.tex}

\section{Open problems}
\input{open.tex}

\input{acknow.tex}

\bibliographystyle{prsty}
\bibliography{co2a}

\end{document}

%% file: introduction.tex
Complex time dependence in laboratory systems, in our natural
environment or in living beings can have a variety of origins.
One of the most fascinating perspectives is represented by the 
description of aperiodic fluctuations in terms of deterministic
dynamical models.
In the last two decades, much work has been devoted to test for this 
hypothesis and to characterise the underlying dynamics under the
assumption that only a scalar time-series is available. Since the 
pioneering articles of Packard et al.\cite{Packard80}, Takens \cite{Takens80}, 
and Grassberger \& Procaccia\cite{Grassberger83d}, a sound body of knowledge 
has been progressively acquired \cite{Kantz97c}, leading to the establishment 
of a new discipline, the nonlinear time series analysis. The general approach
consists in reconstructing the phase space from the observed scalar data, 
most often by making use of the time delay embedding. In a sequence of 
spaces of increasing dimension, one looks for the manifestation of 
deterministic structures such as finite attractor dimension or enhanced 
predictability. Unfortunately, this approach suffers from severe 
limitations as soon as the dynamical complexity of the underlying dynamics 
becomes relatively large.

Systems with time delayed feedback can create arbitrarily complex
dynamics already with very few variables and rather simple equations of
motion. The Mackey-Glass equation\cite{Mackey77} is the best known such 
example. It is a first order scalar differential equation with a force field
that depends on a past value of the variable itself. This model was suggested
in a physiological context (regulation of the production of red blood cells), 
where the mechanism of time delayed feedback is rather common.  Further 
examples range from such widespread scientific disciplines as biology, 
epidemology, physiology, or control theory \cite{Hale77,Bellmann63}.
In physics, this class of systems has been largely ignored, although
time delayed feedback has been introduced in several laboratory experiments 
as an additional means to enhance chaotic properties of systems, as e.g.\ 
in the CO$_2$ laser experiment performed in Ref.~\cite{Arecchi86}. From the 
mathematical point of view, time delayed feedback leads to delay-differential 
equations (see \cite{Hale77} for some results about the existence and 
uniqueness of solutions of the initial value problem). The corresponding
phase space is infinite dimensional, as the initial condition is a generic
function defined on the interval $[-\tau_0,0]$, with $\tau_0$ being the 
delay time of the feedback loop. In practice, however, high frequency 
components are almost absent and thus a finite number of variables suffices
to parametrize the asymptotic solutions. On the other hand, the fractal 
dimension $D$ can be made arbitrarily large as it has been established that
$D$ is proportional to $\tau_0$ for sufficiently large $\tau_0$ 
\cite{Farmer82,GiaLePo95}.

As already mentioned, the direct reconstruction of attractors from scalar 
data through time delay embedding using Takens theorem 
is clearly limited to low dimensional 
objects. A recent estimate \cite{Olbrich97} which takes entropy-related 
folding effects of the embedding procedure into account, shows that the 
minimal number of points $N$ required for a clear manifestation of 
determinism must be larger than $\sqrt{{\rm e}^{hD}s^D}$, where $s$ is the 
required scaling range (e.g.\ $s=10$ represents one decade of scaling) and
$h$ is the Kolmogorov-Sinai entropy. In 
practice, attractors with dimensions larger than 5 can hardly be identified 
by time series analysis using Takens theorem, since 
otherwise an unrealistic large amount of data and an 
unrealistically low noise-level would be required.  

High dimensional attractors of systems with time delayed feedback are thus 
practically indistinguishable from colored noise. On the other hand, the 
underlying delay-differential equation couples only a few variables, so that 
it is natural to ask whether more effective techniques exist, which are able 
to reproduce the observed dynamics. It turns out that a reconstruction, not 
of the attractor in a proper phase-space, but of the dynamical rule in what we
call ``state'' space, is often easier and equally effective. On that basis, 
the delay times of unknown scalar systems 
from a time series with the help of appropriate indicators 
\cite{Fowler93,Tian97} were estimated. 
Later, it was shown that the dynamical rule itself 
can be reconstructed from the time series of scalar time delay systems
\cite{Buenner96,BuennerPRE96a,Voss97,Ellner97} and thereby the Lyapunov 
spectrum \cite{Hegger99}.
Most importantly, the dimension of the state space does not depend on the
delay time, opening up the possibility to model and characterize 
high-dimensional regimes as well.

Since the restriction to scalar time-delay systems is, in practice, too 
severe, some efforts have been made to extend the latter ideas to the case 
of multi-variate time-delay systems. On a phenomenological basis, it was 
demonstrated that the delay time can be estimated also in such systems
by treating the system analogously to a scalar one \cite{BuennerEPL98}.
For a multi-variate delay system with a single time-delayed feedback, an
embedding-like theorem for delay systems was derived and applied to 
experimental data from a laser \cite{Hegger98}. 
For this end, an extension of Takens theorem to input-output systems
as conjectured by Casdagli \cite{Sauer91a,Casdagli92} and later 
proven in \cite{SBDH97}, was applied to time delayed systems.
For the general case of 
multi-variate delay systems (with multi-variate delays) until now a 
multi-variate measurement is required \cite{BuennerPRE97b}. 

In this paper, we discuss in depth the theoretical aspects of the 
identification of a suitable state space for time delay feedback systems. We 
shall first consider the problem of mapping the original dynamics (possibly
characterized by several variables) onto scalar models under the only 
restriction of a single feedback process. In the first part of Sec. II, the 
discussion is carried on for discrete-time models. The result is then
extended to continuous-time models.
In particular, we show that the reconstruction is possible both when the 
recorded variable is and when it is not the feedback variable. The only 
(important) difference between the two cases concerns the minimum dimension
 of the state-space such that a faithful reconstruction is possible: 
The minimum dimension  turns out to be definitely larger in the latter case. 
In Sec. III, we 
discuss the approximations involved in the modellization of continuous-time 
dynamics in terms of delayed maps.

A thorough discussion of the various difficulties encountered in the
practical implementation of these theoretical ideas is then presented in 
Sec. IV with reference to the generalized Mackey-Glass model: a
differential delay equation involving two variables. Problems like the 
determination of the delay-time and the intrinsic limitations of local 
indicators are investigated therein. Sec. V is devoted to the discussion of 
global indicators, while the open problems are briefly reviewed in Sec. VI. 
In the second part of this paper, these concepts will be employed and 
illustrated in the case of experimental data taken from a CO$_2$ laser with 
feedback.

%% file: model1.tex
In this subsection, we introduce multicomponent systems with
delayed feedback and discuss the possibility to map them onto suitable
scalar models. Besides addressing a general mathematical question
(i.e. the equivalence between different classes of dynamical equations),
our motivation resides in the possibility to reconstruct the dynamics
of a delayed system from a single scalar variable.

As anticipated in the introduction, we shall refer to a general case with
$d$ variables. The only restriction that we impose concerns the
number of feedback processes: we shall assume that only one variable is fed
back. We believe that this is a sufficiently general standpoint to begin a
meaningful study of delayed systems.

Although the physically meaningful models are continuous-time systems,
it is worth considering also delayed maps (DM), since the way DDEs are
implemented on digital computers is precisely by constructing a suitable DM
and, more important, DMs can be studied more efficiently to extract the
relevant physical properties from experimental signals. More precisely, we
shall also consider the generic $d$-component DM
\be
\vec{y}(n+1) = \vec{F}(\vec{y}(n),y_1(n-\tau_0)),
\label{eq.dm}
\ee
where $\vec{y}_n \in {\cal R}^{d}$ and the delay time $\tau_0$ is a positive
integer number. The initial condition of the DM consists of a
$(d + \tau_0)$-component vector, so that the phase space is 
${\cal R}^{d + \tau_0}$.
Again without loss of generality, the  feedback variable is assumed
to be the first component.

With reference to discrete-time systems, we now discuss the question
of reconstructing the dynamics of a given component in terms of the
values of the same component at different times. The embedding
theorems~\cite{Takens80,Sauer91a} tell us that the knowledge of
sufficiently many values of $y_k(n)$ for a given $k$ (the chosen
component) suffices to reconstruct the dynamics on the attractor.
More precisely, it is possible to express the value of $y_k(n+1)$ as a
function of its $2D$ previous values, where $D$ is the attractor
dimension. The point we want to address here is the possibility to
reconstruct the dynamics with much less variables than required by the
embedding theorem.

We start from the simple assumption of a linear dynamical system
\be
\vec{y}(n+1) =  {\bf A} \vec{y}(n) + \vec{\alpha} y_1(n-\tau_0)
\label{eq.linear1}
\ee
where ${\bf A}$ is a $d \times d$ matrix and $\vec{\alpha}$
is a $d$-component vector. Next we need to specify which variable is
actually recorded; the structure itself of the above equation reveals a
difference between the first variable (the only one being fed-back) and
all the others. We will see that such a difference plays an important role
in the construction of an optimal model.

We first consider the case of the variable $y_1$ being recorded. The
problem we want to discuss is that of finding the minimum amount of information
to determine $y_1(n+1)$, when the only available information consists in
the past values of $y_1$ itself. All 
components of $\vec y$ are, in principle, necessary but we shall see that they
are implicitely determined so as to make possible a truly
deterministic reconstruction of the dynamical rule only from the knowledge
of $y_1$. Therefore, we define all components of $\vec y$ except $y_1$
as ``hidden variables'', i.e. unknowns that must be determined in order
to be explicitely eliminated from the final dynamical law.

\begin{center}
\begin{figure}
\psfig{file=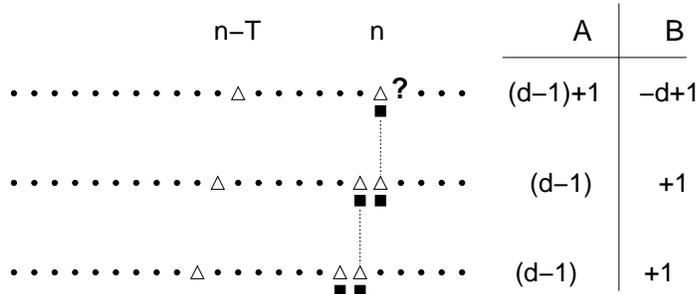,width=10cm,angle=270}
\caption{Illustration of the coupling of the variables in the case
where the embedding aims at the elimination of all the variables except
the one with feedback. Full squares denote the hidden variables. Open
triangles denote the the variable with the time delayed feedback, which
is accessible to measurement.}
\label{fig.rec1}
\end{figure} 
\end{center}

We will now discuss the problem of the information needed to construct a
model in a pictorial way, by referring to Fig.~\ref{fig.rec1}. We hope that
this will be clear enough to be easily followed without the need to enter
technicalities. Each row in Fig.~\ref{fig.rec1} is a schematic description
of the information involved in the application of Eq.~(\ref{eq.linear1}) at a
specific time. A full square positioned in the site $n$ of the
time-lattice indicates that all the $d-1$ hidden variables  at time
$n$ are required in the iteration of the DM. Let us start from the uppermost
row. As we have to determine only $y_1$ at time $n+1$ (see the question
mark in the figure) we need to consider only one equation which, in 
general will however
depend on all variables at time $n$ (see the corresponding  full
square at time $n$) and on $y_1$ at time $n-\tau_0$ (see the triangle). As a
result, we have $(d-1)$ (the full square) plus 1 (the question mark)
unknowns. This number is reported in column $A$ on the right of the figure.
The net difference between the number of available equations and that of
unknowns is instead reported in column $B$. We see that, in this case, since
we have considered just one equation, such a difference is precisely $1-d$.
Accordingly we reach the trivial conclusion that we cannot determine
$y_1(n+1)$ from the knowledge of only $y_1(n)$ and $y_1(n-\tau_0)$.

Further information can be obtained from the past history. Let us
start discussing the application of the dynamical law at the previous
time step (see the second row in Fig ~\ref{fig.rec1}). 
In this case, all the $d$ equations enter into play. 
The number of unknown
variables is $(d-1)$, i.e. the square at time $n-1$, while the
difference between new equations ($d$) and new unknowns is
1. Therefore, the addition of this new step allows reducing the global
gap between unknowns and equations.  Iterating this procedure $d-1$
times will eventually allow to reach a break-even point, when the
number of equations is equal to the number of unknowns. This means
that we have to consider $d$ rows in Fig.~\ref{fig.rec1}, i.e. that
$y_1(n+1)$ is unambiguously determined once $y_1$ itself is known in
at least two windows of length $d$. Formally, we find that for $m \ge
m_d$~\footnote{For the remainder of this paper we will term $m_d$ the
minimal ''window size'' of the model (\ref{eq.dmr})
that guarantees a proper embedding.}, with $m_d = d$, 
\be
 y_1(n+1) = {\vec \beta} {\vec v}(n;m,\tau_0) ,
\label{eq.dmr}
\ee
where
\bea
{\vec v}(n;m,\tau_0) =  &\bigg(& y_1(n),y_1(n-1),\ldots,y_1(n-m+1); 
\nonumber \\
& &y_1(n-\tau_0),y_1(n-\tau_0-1), \ldots,y_1(n-\tau_0-m+1) \bigg) ,
\eea
an expression stating that we have been able to transform the initial
multicomponent  DM (\ref{eq.linear1}) 
into a scalar equation (\ref{eq.dmr}). The price we had to pay is that now
the dependence on the past values of $y_1$ is not restricted to a single
value as originally assumed in Eq.~(\ref{eq.linear1}), 
but $d$ consecutive values are needed.

If $\tau_0<d$, the $2d$ variables appearing on the l.h.s. of Eq.~(\ref{eq.dmr})
overdetermine $y_1(n+1)$, since in the above described process some unknowns
are counted twice. As we have in mind applications to models with a few
components compared to the delay, we shall not argue further about this
point. Moreover, it is instructive to see that the dimension of the
phase-space is $\tau_0+d$ in the reconstructed model as well as in the original
one: Iteration of Eq.~(\ref{eq.dmr}) indeed requires knowing $y_1(l)$ in the
whole range $n \ge l > n-\tau_0-d$. Accordingly, model Eq.~(\ref{eq.dmr})
provides a faithful reconstruction of the whole dynamics including the
convergence to the asymptotic attractor. This is to be contrasted with
the possibility offered by the standard embedding technique to describe
only the dynamics on the attractor itself.

The advantage over the standard application of the embedding theorem
becomes more transparent if we also notice that the number of variables
needed to reconstruct the dynamics is $2d$, independently of the delay
$\tau_0$, i.e. independently of the phase-space dimension $\tau_0+d$ that 
can be arbitrarily large. In particular, the technique can be equally effective
also in the high dimensional regimes generally existing whenever
$\tau_0 \gg 1$
(let us recall that the dimension of the attractor is proportional to
the delay).

In the case of a nonlinear DM (\ref{eq.dm}), the basic difference is that the
function $\vec{F}$ is, in general, non-invertible. This implies that
longer sequences of variables must be considered to remove the ambiguities
inherent to the lack of invertibility. In analogy to the embedding theorem,
it is natural to conjecture that the model equation
\be
 y_1(n+1) = f( {\vec v}(n;m,\tau_0)) ,
\label{eq.dmnr1}
\ee
with two windows of length $m \ge m_d$, and $d \le m_d \le 2d+1$, 
suffices to faithfully reconstruct the dynamics even in the worst case.
This conjecture is indeed confirmed, if we interpret the delayed feedback as
an ``external'' driving and thus see the whole system as an
input-output system like those considered by \cite{Casdagli92}. This
analogy, suggested in \cite{Hegger98}, allows referring to the
generalization of embedding theorems reported in Ref.~\cite{SBDH97}, which
precisely indicate that $2(2d+1)$ is a true upper bound for the number of
variables to be actually used in the model reconstruction. 

\begin{center}
\begin{figure}
\psfig{file=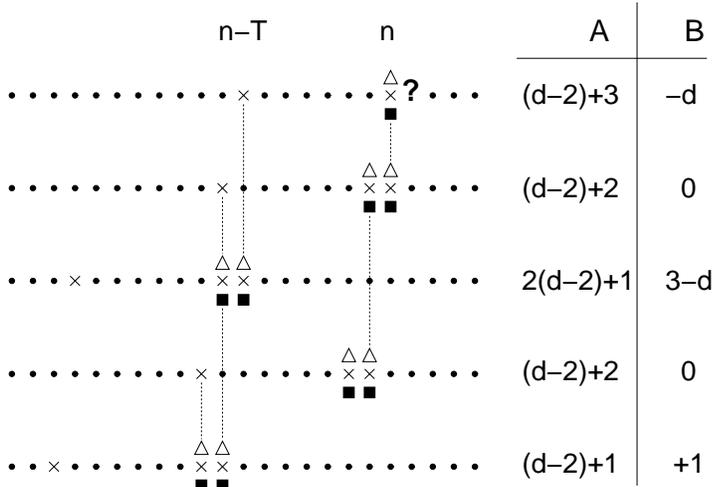,width=10cm,angle=270}
\caption{Illustration of the coupling of the variables in the case where
the embedding aims at the elimination of the variable with feedback. Hidden
variables without feedback are denoted by full squares. The hidden variable
with feedback is denoted by crosses. The experimentally observed variable 
is denoted by open triangles.}
\label{fig.rec2}
\end{figure} 
\end{center}

The feedback variable is certainly peculiar and different from all other
variables involved in the dynamical process. It is therefore, interesting to
ask oneself whether a compact reconstruction of the model is still possible
if {\em not} the feedback variable is measured, but any other variable. 
The answer is yes, but the number of variables to
acquire the necessary information is larger than before and the proof is
also rather cumbersome so that a pictorial represention such as the one
reported in Fig.~\ref{fig.rec2} will be very helpful. In this
case, we must distinguish among three types of variables: $(d-2)$ hidden
variables without feedback (represented by a full square); 
the hidden feed-back variable (cross)
and the variable experimentally observed (open triangle). In the first step
of the procedure, there is one more unknown variable than before (since $y_1(n-\tau_0)$
is unknown, too) so that the gap between variables and unknowns is $-d$.
Equally more negative is the second step, since the existence of an
additional variable (the feed-back which is not recorded) prevents having a
net gain. Therefore, recursively repeating the very same step does not allow 
removing all the unknowns. Nevertheless, we can still
find a meaningful solution by modifying our strategy as described in the
third step, where we consider the application of the mapping at time
$n-\tau_0-1$. After comparing the newly involved variables with those already
introduced in the two previous time steps, one sees that the additional gap
is equal to $3-d$.
This result is strictly positive only if $d\le2$, thus suggesting that this
new strategy leads in general to worse results. However, from now on,
one can alternate steps of the previous and new type (see. e.g., the fourth
and the fifth line in Fig.~\ref{fig.rec2}): this allows gaining 1 equation
every second step. The break-even point is obtained after $2d-1$ steps. This
means that the recorded variable must be known in two windows of length
$2d-1$. Accordinlgy, the price to be payed for not dealing with the
feed-back variable is that the number of ``variables'' is almost twice as
large as before. Nevertheless, we can still consider this last result as
positive, since the dimension of the space is still independent of the delay.
An important difference with the previous case concerns the
phase-space dimension. The iteration of the reconstructed model
requires now to know a single variable over $\tau_0 + 2d-1$
consecutive times, a number larger than the initial dimension $\tau_0
+d$ if $d>1$. This means that our procedure has enlarged the
phase-space dimension, introducing some spurious directions.  
We want to show now that the price for keeping the dimension of the
phase-space equal to the original value is the construction of a much more 
complicated model. In fact, with reference again to Fig.~\ref{fig.rec2}, we 
see that the steps of type 1 do not allow any gain only until we arrive at 
time $n-\tau_0$. However, from that point the number of unknowns reduces
by unit per single step since the variable $y_1$ was already taken 
into account, so that we eventually do not need to go beyond time
$n-\tau_0-d$. However, in doing so, all variables in the entire delay time
are included, i.e. the standard embedding approach has been followed.

More in general, in the case of nonlinear systems, we expect that a
model 
\be
 y_2(n+1) = f( {\vec v}(n;m,\tau_0)) ,
\label{eq.dmnr5}
\ee
exists for $m \ge m_d, d \le m_d \le 4d-1$.
However, it is honest to recognize that one will be  hardly able to go
beyond $d=2$ in practical cases.

As in the standard embedding technique, the presence of nonlinearities with 
the possible noninvertibility of some functions might require doubling
the number of variables necessary for a faithful reconstruction of the 
dynamics. Note that the number of spurious directions introduced by the
DM-model in the nonlinear case is at most $3d-1$, 
and therefore much smaller compared to the
number of spurious directions introduced by a Takens-type model, which can
be up to $\tau_0+d+1$.

In the case of time-continuous models, we refer to  
first order delay-differential equation (DDE)
of the type
\be
\dot{\vec{x}} = \vec{H}(\vec{x},x_1(t-\tau_0) ),
\label{eq.dde}
\ee
where, without loss of generality, the feedback variable is assumed to
be the first component $x_1$ of the $d$-dimensional vector $\vec x$,
while $\tau_0 \in {\cal R}^+$ is the delay time. The initial condition
for the DDE consists in a differentiable function in the
interval $[t_0-\tau_0,t_0]$ plus a $(d-1)$-dimensional vector
(i.e. the remaining components) at time $t_0$.  Therefore, the phase
space is ${\cal C}^1 [0;\tau_0] \times {\cal R}^{d-1}$.

For time-continuous models (\ref{eq.dde}), basically the same procedure applies, except
that instead of including the dependence on additional times in the past as in 
Eq.~(\ref{eq.dmnr1}), one has to add higher-order time derivatives. The
final results are: 

(1) a scalar DDE equation for the variable $x_1$
of $m$-th order (with $m \ge m_d$
and $d \le m_d \le 2d+1$) 
\be
 x_1^{(m)} = h \left({\vec w}(m,\tau_0)\right) ,
\label{eq.dder}
\ee
where
\bea
{\vec w}(m,\tau_0) =  &\bigg( & x_1,x_1^{(1)},\ldots,x_1^{(m-1)}; \nonumber \\
& & x_1(t-\tau_0),x_1^{(1)}(t-\tau_0),\ldots,x_1^{(m-1)}(t-\tau_0)\bigg) .
\eea
We write $x_1^{(i)}$ for the $i$-th derivative of the variable $x_1$ with
respect to time.

(2) a scalar DDE equation for the variable $x_2$
of $m$-th order (with $m \ge m_d$
and $d \le m_d \le 4d-1)$
\be
 x_2^{(m)} = h \left({\vec w}(m,\tau_0)\right) ,
\label{eq.dder2}
\ee
where
\bea
{\vec w}(m,\tau_0) =  &\bigg( & x_2,x_2^{(1)},\ldots,x_2^{(m-1)}; \nonumber \\
& & x_2(t-\tau_0),x_2^{(1)}(t-\tau_0),\ldots,x_2^{(m-1)}(t-\tau_0)\bigg).
\eea

%% file: model2.tex
In the previous section we have seen that vectorial delay-models can be 
mapped onto scalar ones by embedding the attractors into suitable state spaces 
defined in terms of a single variable recorded in two windows of length $m_d$. 
These results provide the minimal framework for 
reconstructing the dynamics starting from the knowledge of just one 
observable. However, the exact inference of the model is formally possible 
only when a DDE (DM) dynamics is reconstructed in terms of a continuous 
(discrete) time model.

In reality, almost all physically meaningful processes stem from continuous
time equations, while data are typically accessible as sequences of values
sampled with a finite frequency. Accordingly, the typical situation consists
in constructing a DM that mimics a DDE, i.e. we have to deal with
the problem of passing from one to the other class of models. In this section,
we discuss this problem, showing that the model mismatch implies that 
increasingly faithful reconstructions are only possible at the expense of
increasing the state-space dimension. This can be done by lenghtening either
the first or the second window of each pair. 

Before proceeding to the general discussion, it is important to stress
 that all the results reported in this section are derived under the
assumption that (i) there exists a finite-dimensional attractor (this can
be shown under quite general conditions \cite{Hale77}), (ii) the attractor
dynamics is high-dimensional. In particular, we assume that the length of 
the window-pairs is smaller than the minimal embedding dimension required 
by the Takens theorem. This is because, as explained in the introduction, 
we want to consider cases where the usual embedding techniques fail to 
provide a faithful reconstruction.
 
For the sake of simplicity, we first consider a scalar DDE,
$\dot{x}=H(x,x(t-\tau_0))$,
 and assume
that the continuous variable $x(t)$ is recorded with a constant sampling time $\Delta$ 
on the discrete time-lattice $t_0+n\Delta$ with $n \in {\cal Z}$. Let us call
$x(n)=x(t_0+n\Delta)$ from now on. In this
framework we shall investigate the degree of accuracy that is possible to
reach within the class of scalar DM-models.
Let ${\cal A}(m_1,m_2)$ be the class of analytic functions $h$ 
\be
{\cal A}(m_1,m_2) =  \{ h: R^{m_1+m_2} \to R \}.
\label{eq.modelclass}
\ee
Consider the DM-model 
\be 
x(n+1)=h(\vec{v}(n;m_1,m_2,\tau)),
\label{eq.dmnr2}
\ee
with $h \in {\cal A}$, and
\bea 
\vec{v}(n;m_1,m_2,\tau_0)=  & \bigg(& x_1(n),x_1(n-1),\ldots,x_1(n-m_1+1); 
\nonumber \\
& &x_1(n-\tau_0),x_1(n-\tau_0-1), \ldots,x_1(n-\tau_0-m_2+1) \bigg) ,
\label{eq.dmnr3}
\eea
with window pairs $(m_1,m_2)$\footnote{We have introduced the notation
$(m_1,m_2)$ to emphasize that the length of the two windows may be different. In that
respect, the 
definition of $\vec{v}$ contrasts with the one given in the previous section.} 
separated by a time $\tau_0$. 
We quantify the accuracy of the DM-model $h$ in Eq.
(\ref{eq.dmnr2}) with the help of
 the one-step forecast error (FCE):
\be
{\bar \sigma}(h;m_1,m_2,\tau) =\sqrt{ \frac{
       \left \langle \left (
	x(n+1) - h(\vec{v}(n;m_1,m_2,\tau)) \right )^2 \right \rangle }
       {\langle x(n)^2 \rangle - \langle x(n) \rangle^2}}
\label{eq:gfce}
\ee
where $\langle \cdot \rangle$ denotes a time average. 

Any model $h$ can be geometrically seen as an $(m_1+m_2)$-dimensional 
manifold in the state space augmented by the $y(n+1)$ direction (we shall
call it, the $\cal S$-space). The FCE is trivially larger than zero
whenever the original data lie on a manifold different from that one
identified by the model. This is an error that can be removed by properly 
constructing the model. Conversely, if the data are distributed in a broader 
region, i.e. also transversally with respect to a hypothetical manifold,
no exact model can be constructed and the FCE is bounded away from zero. 
This is precisely what we expect to happen because of the model mismatch:
for any choice of window $(m_1,m_2)$, the variable $y(n+1)$ fluctuates  
in a small but finite interval, so that the FCE cannot be smaller than 
the average thickness of the distribution of points.

In order to clearly distinguish the latter fundamental limitation from
trivial modelling errors, it is sufficient to define 
${\bar \sigma}_{\cal A}(m_1,m_2,\tau)$ as
\be
{\bar \sigma}_{\cal A}(m_1,m_2,\tau)  = min \{ {\bar \sigma}(h;m_1,m_2,\tau) | h \in 
  {\cal A}(m_1,m_2,\tau) \}.
\label{eq.fcemin}
\ee
${\bar \sigma}_{\cal A}(m_1,m_2,\tau)$ establishes the 
maximum level of accuracy that can be reached 
with a fixed window-system $(m_1,m_2)$ and a delay time $\tau$ 
in the class of DM-models ${\cal A}$. From now on the
function $h \in {\cal A}$, which minimizes the FCE, is 
called $\hat{h}$ and therefore 
${\bar \sigma}_{\cal A}(m_1,m_2,\tau) ={\bar \sigma}(\hat{h};m_1,m_2,\tau)$.
We shall see that there are at least two 
alternative procedures to increase the accuracy of the reconstruction. 
They amount to considering window pairs of the type $(1,m_2)$ and $(m_1,1)$, 
respectively.

Let us first discuss the $(1,m_2)$ case. Uniqueness and existence 
theorems \cite{Hale77} guarantee that the original DDE model can be written 
as a functional
\be
 x(t+\Delta) = G[x(t),\{x\}_d]
\label{eq.fdde}
\ee
where $\{x\}_d=\{x(t')|t-\tau_0 \le t' \le t-\tau_0 +\Delta\}$. 
A simple example of the above functional dependence can be obtained in the
case of the model class
\be
  \dot x = -\mu x + F(x(t-\tau_0))
\label{eq.img}
\ee
to which both the Ikeda \cite{Ikeda87} and Mackey-Glass \cite{Mackey77} models 
belong.  A formal integration of Eq.~(\ref{eq.img}) yields
\be
x(t+\Delta ) = x(t){\rm e}^{-\mu \Delta} + \int_0^\Delta dt' F(x(t-\tau_0+t')) 
{\rm e}^{\mu (t' -\Delta)}
\label{eq.int1}
\ee
If $x(t-\tau_0+t')$ is nearly constant within the integration
interval, one can approximate the functional dependence with a single
value of the variable $x(t-\tau_0+t')$ within the integration
interval. This amounts to constructing a $(1,1)$-model and the 
uncertainty on $x(t+\Delta)$ is precisely the above introduced FCE 
${\bar \sigma}_{\cal A}(1,1,\tau_0)$, which is of $\Delta^2$-order. \footnote{In this section
we always assume that the delay $\tau_0$ is perfectly known and the
uncertainty is entirely due to a model mismatch.} 

A better accuracy can be achieved if two or more consecutive points
are assumed to be known in the vicinity of $x(t-\tau_0)$, since their
knowledge allows constructing higher order approximations of $F$.
Simple perturbative arguments suggest that the error made in the estimation of
$x(t+\Delta)$ is of the order $\Delta^{m_2+1}$, if $m_2$ consecutive points are
used (i.e., if a window-pair $(1,m_2)$ is considered) and $\Delta$ is
small enough. In fact, the problem of estimating the error for fixed
$\Delta$ and $m_2$ large enough is absolutely non trivial and deserves a
discussion by its own. Here, without pretending to derive asymptotic estimates
on the dependence of ${\bar \sigma}_{\cal A}(1,m_2,\tau_0)$ on 
$m_2$ and $\Delta$, we limit
ourselves to consider two limit cases. The first one consists in assuming 
that the Fourier modes above a certain frequency $\omega_c$ are slaved modes, 
i.e. they are uniquely determined by the amplitude of the lower-frequency 
modes. In this case, if the sampling time $\Delta < 2\pi/\omega_c$, we 
expect the residual uncertainty on $x(t+\Delta)$ to vanish for increasing 
$m_2$, although it is not obvious to determine how rapidly.  
In the opposite limit, we can assume that the amplitude of each 
high-frequency mode is an independent variable (as in a stochastic process). 
In this case, the uncertainty on $x(t+\Delta)$ would depend on the ``power'' 
contained in the Fourier spectrum above the sampling frequency 
$\omega_s = 2\pi/\Delta$ and would not decrease for increasing $m_2$. Were
this the typical condition generated by DDEs, one should conclude that the 
model mismatch is so severe that one can never reproduce a continuous-time
dynamics with arbitrary accuracy (with the exception of $m_2$ larger than
the minimal dimension required by the embedding theorems for a faithful
reconstruction).

An alternative approach for constructing a DM consists in approximating
the first-order time derivative with linear combinations of the observable 
$x$ in neighbouring points along the time lattice. It is well known that 
one can write
\be
 \dot x\left(t-(\frac{m_1}{2} -1 )\Delta\right) = 
 \frac{1}{\Delta}\sum_{i=-m_1+1}^{1} a_i x(t+i\Delta) + {\cal O}(\Delta^{m_1+2})
\ee 
for a suitable choice of the coefficients $a_i$. Upon substituting the
above expression in the initial DDE and solving for $x(t+\Delta)$, we
find that $x(t+\Delta)$ can be expressed as a function of the $m$ preceding
values and 1 value one-delay unit back in time. In other words, we have
arrived at a DM of type $(m_1,1)$, which involves an unavoidable error
${\bar \sigma}_{\cal A}(m_1,1,\tau_0) \simeq \Delta^{m_1+1}$. This is again a purely 
perturbative result which is valid only for moderately large $m$'s.

In both the above discussed cases, we have seen that a discrete-time model 
can reproduce only approximately the dynamics of the original 
continuous-time system. In comparison to low-dimensional dynamical systems, 
for which we know that a generic ODE can be exactly transformed into a discrete
mapping (even with the additional advantage of reducing the phase space
dimension, if a Poincar\'e section is taken), the above
results look very modest. The main reason for such a difference is
that when a DDE is turned into a DM, the phase-space is necessarily
``compressed'' from an infinite- to a finite-dimensional one. The
compression may be practically harmless, but necessarily involves the
loss of small but nonzero interaction terms.

The two pairs $(1,m_2)$ and $(m_1,1)$ are the limit cases of the more
general combination $(m_1,m_2)$. We have been unable to estimate directly
the uncertainty in this general case, because we failed to find an
interpretation of the  corresponding model in terms of derivatives and/or
integrals. Nevertheless, with the help of a recursive argument we conjecture
that 
\be
   {\bar \sigma}_{\cal A}(m_1,m_2,\tau_0) \simeq \Delta^{m_1+m_2} .
\label{eq:generr}
\ee
We show this by starting from a DM model of the type $(m_1,m_2)$, namely 
\be
x(n+1) = F^{(1)}(\vec{v}(n;m_1,m_2,\tau_0))
\label{eq:mod1}
\ee
which we assume to be accurate up to order $\Delta^{m_1+m_2}$. Moreover, we
can claim that, as long as the distance between the two windows of
a given pair is $\Delta$-close to the true delay $\tau_0$, the error of the 
corresponding model does not change significantly. Accordingly, the 
dynamics is equally well described by the model
\be
x(n+1) = F^{(2)}(\vec{v}(n;m_1,m_2,\tau_0-1)).
\label{eq:mod2}
\ee
By now solving this latter equation with respect to $x(n+1-m_1)$ 
(assuming that no problems connected with the invertibility of the 
nonlinear expression arise) and substituting in Eq.~(\ref{eq:mod1}), we can
write 
\be
x(n+1) = F^{(3)}(\vec{v}(n;m_1-1,m_2+1,\tau_0))
\ee
for some function $F^{(3)}$. In other words, the value of
$x$ at time $n+1$ can be predicted on the basis of a window-pair of the type
$(m_1-1,m_2+1)$ with the same order of magnitude for the uncertainty,
i.e. $\Delta^{m_1+m_2}$ (in fact, the additional factor due to the error
propagation in the inversion of the second equation is a finite correction
term, independent of $\Delta$). By iterating the same argument, one can 
eventually convince oneself that the accuracy of the model depends only on 
the total number of points in the two windows.

Finally, we briefly discuss the general case of how to approximate 
multicomponent DDEs with multicomponent DM- models. In
order to avoid technicalities, we limit ourselves to summarize the
main steps, the whole derivation being straightforward. The
generalization of the first approach (leading to (1,m)-models in the 
scalar case) confirms the naive expectation
based on the knowledge of the scalar case, i.e., window pairs of the 
type $(m_d,m_d+l)$ guarantee an error of the order $\Delta^{l+2}$.
In fact, in full analogy with the discussion of Eq.~(\ref{eq.int1})), one 
can conclude that the formal integration of all equations allows approximating
the original DDE equation up to order $\Delta^{l+2}$ with a ``generalized'' 
DM, where $l+1$ past values of the scalar feedback variable are required.
A simple repetition of the arguments presented in the first part of this 
section shows that this vector map can be turned into a scalar one of the
type $(m_d,m_d+l)$.

In the complementary case that has led to the development of (m,1)-models in
the scalar case, the scenario is much worse, since the derivative of each of 
the $d$ variables must be determined with the prescribed level of accuracy. 
In order to fulfil this requirement, one must transform the original DDE into
a DM involving $dm_d$ variables in the first window. A by far larger number 
of variables is required as soon as $d>1$.

%% file: implementation.tex
The most general delayed systems involve the dynamics of several components. 
In practice, however, only a single scalar variable is available.
Hence, it is natural to reconstruct the dynamics in terms of intrinsically 
discrete models such as DMs. This choice of model class is further 
motivated by the numerical instabilities that are known to affect the 
computation of derivatives (required in the practical implementation of 
DDE models).

There are several ways to quantify the deviations from the expected dynamics 
in delayed systems, such as the filling factor \cite{Buenner96,BuennerPRE96a}, 
the ACE-method \cite{Voss97}, and others \cite{Ellner97}. 
 For the sake of 
simplicity, here we restrict our investigations to $(m,m)$-DMs,
\be
y(n+1)=h(\vec{v}(n;m,\tau)).
\label{eq.model1}
\ee
In the following, 
we shall use the one-step forecast error $\sigma(h;m,\tau)={\bar \sigma}(h;m,m,\tau)$,
and its minimum in the set ${\cal M}$:  
$\sigma_{\cal M}(m,\tau)=\sigma(\hat{h};m,\tau)$, where the \
function $\hat{h} \in {\cal M}$ minimizes $\sigma(h;m,\tau)$,
as a tool both to 
identify the correct delay time and to construct a meaningful model.
In practice, one cannot deal with such a large space like that of analytic 
functions ${\cal A}$ considered in the previous section. Accordingly, one must first 
identify a proper class of parametrized functions $h$ to work with:
\be
{\cal M}(m) =  \{ h(\cdot;\vec{a}): R^{2m} \to R \},
\label{modelclass}
\ee
where the parameter $\vec{a}$ is varied to minimize the FCE.  
The optimal choice of a specific class ${\cal M}(m)$ depends on the problem 
under consideration. In practice, however, local linear models or global 
models built by radial basis functions~\cite{He93a} are generally
quite successful.  Here, we stick to the former class. The average required
by the definition (\ref{eq:gfce}) is obviously performed along the available 
time series.

In practice, besides the fundamental limitations discussed in the previous
section, several additional factors like finite sampling time $\Delta$, 
measurement noise, finite number $L$ of data, mismatch between the delay 
time $\tau_0$ and the actual sampling time, i.e. 
$mod(\tau_0,\Delta) \neq 0$ (note that so far in the literature only the 
case of no mismatch has been discussed). While the effect of noise will
be considered in the second part of the paper with reference to a truly 
experimental system, here we shall investigate whether the other limitations
may actually obstruct the model reconstruction. 

The approach adopted in this section consists in identifying the optimal 
model $\hat{h}$ in ${\cal M}(m)$ as the one minimizing $\sigma$ 
(as in the previous section) and then finding the minimum value of $m$ and 
the appropriate value of the delay $\tau$ such that the ``distance'' 
$\sigma_{\cal M}(m,\tau)$ of the reconstructed model (\ref{eq.model1}) from the 
true dynamics is sufficiently small.

However, it is important to notice that local closeness between the model and
the true dynamics does not necessarily imply closeness of the global dynamics. 
We can see this by discussing the case of a grossly wrong $\tau$-value 
wherein we can expect that $y(n+1)$ is almost totally uncorrelated with the 
$y$-values belonging to the second (delayed) window. Accordingly, the 
information content of the second window is totally irrelevant in the 
minimization procedure of FCE. By invoking, as in the previous section, the 
analyticity properties of the underlying signal $y(t)$, we can estimate that
$\sigma_{\cal M} \simeq \Delta^{m}$ (since knowing the value of $y$ in $m$ points 
is tantamount to knowing the first $m$ derivatives). This means that the 
FCE can be made very small even in the absence of relevant information 
about the force field (we recall that the delay is assumed to be far from
the correct value), a conclusion that appears utterly illogical. In fact, this 
result tells us that in the small sampling-time limit, it is possible to 
perform reasonable short term predictions by simply exploiting the smoothness 
of the signal itself. In particular, it is not even possible to distinguish 
the true dynamics from that of the naive model $y^{(m)} = 0$, 
which corresponds to polynomial dependence on time, i.e. a dynamics which 
is neither stationary nor even limited. The conclusion to be drawn from 
this observation is that the smallness of $\sigma_{\cal M}$ alone is not enough to 
conclude that a meaningful model can be extracted from the raw data.
This is the reason why we devote the next section to the discussion of other, 
global, indicators which do not suffer the same problems.

For $m_d=1$ (scalar DDEs), this proved to be a very effective and numerically 
inexpensive strategy to detect the unknown delay time $\tau_0$ from time 
series, since $\sigma_{\cal M}$ displays a pronounced local minimum for
$\tau=\tau_0$ \cite{Hegger98}. Before presenting the numerical data, let
complete the general discussion about the FCE by comparing the previous
considerations with the expection for $\tau=\tau_0$. From the discussion 
carried on in the previous section, if $m \geq m_d$, the FCE is at least 
of the order of $\Delta^{2(m-m_d+1)}$. By comparing this estimate, derived 
for the correct value of $\tau$, with the typical error expected for a 
generic delay, we find that the latter one is smaller, if $m < 2(m_d-1)$, 
which is a clear nonsense. Since the FCE is defined as the
{\it optimal} error, whenever some prior information is given, we can conclude
that whenever both mechanisms do apply (i.e. when $\tau = \tau_0$), it is 
the most efficient one which determines the actual FCE. In other words,
we do not expect any sensible dependence of $\sigma_{\cal M}(m,\tau)$ on $\tau$ if
$m < 2(m_d-1)$ (preventing the detection of the delay time with the help of
the FCE), while a clear minimum should be seen in the opposite case.
We can explain that behavior of the FCE  by 
noticing that the two estimates of $\sigma_{\cal M}$ have been derived by invoking 
different mechanisms: (1) continuity of the evolution, (2)  effective 
approximation of the delayed feedback model. Since both mechanisms allow for high-quality
short-term predictions, both will lower the FCE (and supposedly any local
indicator). Therefore, local indicators are not appropriate tools to 
distinguish between the the two mechanisms. Global indicators (as discussed
in the next section) are good candidates to also detect the delay times
for  $m < 2(m_d-1)$.
Anyway, the above inequality represents a necessary condition to be
satisfied by a local indicator (such as the FCE) for a correct identification
of the delay in the worst possible case.

In the following we discuss the problem of model reconstruction with reference
to the generalized Mackey-Glass system \cite{BuennerPRE97b}
\begin{eqnarray}
\dot{y}(t)&=&\frac{ay(t-\tau_0)}{1+y^{10}(t-\tau_0)}+x(t) \label{eq.gen_mkg},\\
\dot{x}(t)&=&-\omega^2y(t)-\rho x(t). \nonumber
\end{eqnarray}
which can be easily transformed into a second order ($d=2$), neutral DDE 
\begin{equation}
\ddot{y}(t)=-\omega^2 y(t)-\rho\dot{y}(t)+\omega^2f(y(t-\tau_0))
 + \frac{df(y(t-\tau_0))}{dy(t-\tau_0)}\dot{y}(t-\tau_0).
\label{general_mkg2}
\end{equation}
The parameters are chosen as $a=3.0, \rho=1.5, \omega =1.0$, and
$\tau_0 =9.83$, for which the Kaplan-Yorke dimension of
the attractor is $D_{KY}= 7.2$. For $\rho = \omega^2$ and $\rho \to \infty$, 
the above equation reduces to the standard Mackey-Glass system by 
eliminating adiabatically the variable $x(t)$.

For the analysis, we use a time series of the variable that is fed back, 
$y(t)$, with a sampling time $\Delta=0.1$. Notice that with this choice of
$\Delta$, the retarded values $y(t-\tau_0)$ lie outside the time lattice 
if $y(t)$ corresponds to one of the sampled values. In 
Fig.~\ref{fig.timeseries_gmkg} portions of the time series and a delay 
plot of an extremal section $\dot{y}(t_i)=0$ are presented. The effect of 
the second component $x$ in the dynamical equation (\ref{eq.gen_mkg}) can 
be clearly visualized in the delay plot (with the delay being close to the
delay time $\tau_0$), since the intersection points of a scalar system
have to lie on curve in such a representation \cite{Buenner96}.  

\begin{figure}
\caption{(a-b) Time series of the generalized Mackey-Glass system;
(c) delay plot of an extremal section. The values of extremal points
$y(t_i), \dot{y}(t_i)=0$ are plotted versus its retarded values.}
\label{fig.timeseries_gmkg}
\end{figure} 

\begin{figure}
\psfig{file=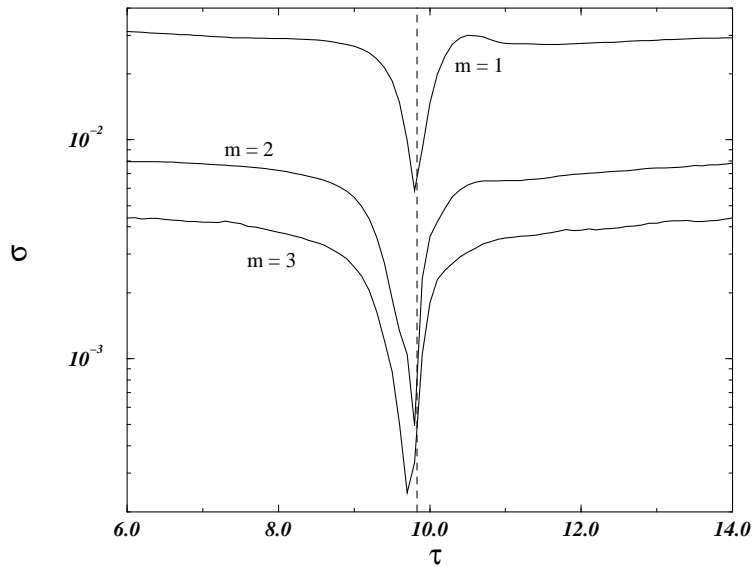,width=10cm,angle=270}
\caption{One-step forecast error of an $(m,m)$-model as function of $\tau$.
From top to bottom, the curves refer to $m=1$, 2, and 3, respectively.}
\label{fig.forec_gmkg}
\end{figure}

We have numerically investigated the FCE $\sigma_{\cal M}(m,\tau)$ 
with a local linear model for different
choices of $m$ and $\tau$.  The FCE is computed taking into account 
time series of length $L(m=1)=50,000;L(m=2)=100,000;L(m=3)=200,000$.
From the data reported in Fig.~\ref{fig.forec_gmkg}, it is interesting to
notice that a pronounced minimum of $\sigma_{\cal M}$ is observed even for $m=1$,
when, a priori, there is no reason to expect a faithful reproduction of the
original dynamics. Such a result is the consequence of a general feature
of dissipative systems: the various components are not equally ``active''.
Indeed, as long as the attractor is highly dimensional, the feedback term
can be viewed as a noise term. In the absence of this ``noise'' source
\cite{Dorizzi87}, 
the original system reduces to an ordinary differential equation, whose
attractor fills a manifold of dimension smaller than $m_d$. The addition of 
the noise ``thickens'' the distribution along all directions in the state
space, the width of the distribution  depending on both 
transverse stability and noise amplitude. 
Accordingly, it may happen that
the role of some components (corresponding to rather stable directions) 
in a multidimensional DDE
is just to blur the distribution generated by  a suitable DDE with 
less components.
This is, to some extent, what happens in our system as clearly seen in
Fig.~\ref{fig.timeseries_gmkg}(c), where the points cluster around a smooth curve
which is the expected shape for the scalar Mackey-Glass system. 
A $(m=1)$-model will detect this curve, leading to a local minimum in
the FCE.  

The position of the minimum is an estimate $\hat{\tau}(m)$ of the delay time.
A parabolic approximation of the FCE around the minimum yields the data
reported in the following table.
\begin{center}
\begin{tabular}[pos]{c|c|c|c}
  & $m=1$  & $m=2$ &  $m=3$ \\
\hline
\hline
$\hat{\tau}(m)$ & $9.81 \pm 0.1$ & $9.87 \pm 0.1$ & $9.72 \pm 0.1$  \\
\end{tabular}
\end{center}
The estimated values agree with the correct value $\tau_0=9.83$ within
the errors due to the finiteness of the sampling time.

Some comments are in order about the behaviour of the FCE. First of
all, let us notice that a local minimum is observed also for zero
delay. This minimum is due to the fact that we pass from a system of
two windows of length $m$ to a single window of double
length. According to the arguments put forward in the first part of
this section, we have to expect an accuracy of the same order as for the 
leading minimum. However, this accuracy cannot correspond to an equivalent
accuracy of the global dynamics as the information about the feedback is
missing. Moreover, we notice that the plateaus of the various curves 
decrease for increasing $m$. This is qualitatively in agreement with
the considerations presented in the first part of this section about
grossly wrong delays. However, there is no quantitative agreement about
the scaling dependence on $m$: we attribute this to the existence of
residual correlations between $y$ values even when they are some time 
units apart. 

Analogous considerations can be made for the height of the minimum that
decreases less than expected on the basis of the general considerations
discussed in the previous section. In this case, we have identified 
in the accuracy of the local linear model and in the finiteness of the
number of points the main limiting factors which prevent $\sigma_{\cal M}$ from being 
smaller for $m=3$.

%% file: m-estimation.tex
We have seen that the FCE is a useful tool to decide whether a reconstructed 
model $\hat{h}$ is locally close to the observed dynamics. Nevertheless, 
there is neither guarantee that the model dynamics remains confined to the 
region where it has been originally defined (e.g., that it does
not explode) nor that it does not converge to a smaller subset (e.g.,
a fixed point or a limit cycle). In other words, the smallness of
the FCE $\sigma_{\cal M}$ is a necessary but not sufficient condition to establish
whether a given model provides a globally faithful reconstruction. To
test this, we iterate the models $\hat{h}$ for different $m$-values and
the optimal choice of the delay time, $\tau=\hat{\tau_0}(m)$, to
generate some typical time series $\{\hat{y}_m\}$. We consider a model 
as valid, if the resulting attractor is ``close'' to the original one. 
To this aim, we introduce and utilize the cross forecast error, compute
the power spectrum, the probability distribution of the sampled variable,
and the Lyapunov spectrum as tools to establish altogether the validity of 
a given model.

However, before discussing all such indicators, it is instructive to perform
a qualitative analysis of the generalized Mackey-Glass system
(\ref{eq.gen_mkg}) for $m=1,2,3$ and the optimal choice of the delay time 
(as identified in the previous section). The resulting time series 
$\{\hat{y}_m\}$ (of length $L = 100,000$) reveal a qualitative
good agreement with the original series only if $m \ge 2$. Indeed, 
for $m=1$, the time series $\hat{y}_1$ is asymptotically attracted to
either a strictly positive or strictly negative region (see
Fig.~\ref{fig.timeseries_m1})
\begin{figure}
\caption{Iterated time series $\{\hat{y}_m\}$ 
of the $(m=1)$-model: (a) Convergence
to an attractor with purely positive values; (b) blow-up of the 
attractor dynamics; (c) delay plot of an extremal section.}
\label{fig.timeseries_m1}
\end{figure} 
Indeed, attractors with a specific sign exist in the standard Mackey-Glass 
system, where the unstable fixed point $y=0$ acts as an ``impenetrable'' 
domain boundary separating the two coexisting attractors (changes of
sign can exist only if they are present in the initial condition; during
the evolution, once disappeared, they cannot be generated again). 
Since, for $m=1$, the second variable is obviously absent, it is not
surprising that the reconstructed dynamics exhibits typical
features of the standard Mackey-Glass system.

As this eventual convergence towards either positive or negative values
persists, independently of the accuracy used in the model
reconstruction and of the number of data points, we must rule out the 
possibility that $m_d=1$, i.e. that a minimal approximate model can be
constructed with just one component. 

The problem of quantifying the ``closeness'' between the original time 
series $y$ and the iterated time series $\hat{y}_m$ cannot be faced by
measuring to what extent the model-generated time-series remains all the
way close to the original time-series. In fact, because of the chaotic
properties of the evolution, an exponential separation always occurs
which can hide the statistical equivalence of the two time-series. 
The most appropriate approach would consist in definying and measuring 
the distance between the two probability distributions. 
The natural space where this question should be formulated is the 
$(2m+1)$-dimensional space $\cal S$ introduced in Sec. III, i.e. the same
space where the FCE is estimated and the dynamical rule reconstructed.
There are various ways to define a distance, such as the Kullback-Leibler 
information~\cite{BaPo97} or the cross correlation sum~\cite{Kantz94a}. 
Unfortunately, a meaningful implementation is a rather delicate matter. For 
instance, in the case of the Kullback-Leibler information, one needs 
sufficiently many data to get rid of statistical fluctuations in the local 
probabilities. Therefore, we have preferred to introduce a more robust 
geometrical indicator which, although carrying less information, can be
satisfactorily complemented with the implementation of other tools.

For any point $P \in \{ y \}$, determined by following the original 
trajectory, we identify the closest template used to construct the local 
linear model along the iterated time series $\{\hat{y}_m\}$ (after a
suitable transient) and measure the distance $d(n)$ of $P$ from such a 
$2m$-dimensional surface. By averaging the square distances over all 
points $P$, we finally obtain the global indicator
\be
\chi(m) = \sqrt{\frac{1}{N-n_0}\frac{\sum_{n=n_0}^N d^2(n)}
   {\langle y^2\rangle - \langle y \rangle^2} }
\label{eq.xfce}
\ee
where $n_0$ is such that the components of $\vec{v}(n;m,\tau)$ are 
all in $\{y(n)\}$. The definition of $\chi(m)$ is essentially the average 
forecast error along the time series $\{y\}$ on the basis of a model of 
the time series $\{y_m\}$, the latter being restricted to the attractor 
region (cross forecast error~\cite{Schreiber97}).

The results are presented in the following table:
\begin{center}
\begin{tabular}{|c|r|}
\hline
$m$ & $\chi(m)$\\
\hline
1 & 1.197 \\
\hline
2 & 0.032 \\
\hline
3 & 0.026 \\
\hline
\end{tabular}
\end{center}
As a result of the confinement of the dynamics to an attractor with
purely positive values, the distance $\chi(m)$ is large for $m=1$
(actually, so large that it compares with the standard deviation of
the data). For $m \ge 2$, $\chi(m)$ decreases substantially (and could 
be further reduced by increasing the number of data points). Accordingly, 
the minimal choice $m_d=1$ does not yield a faithful reconstruction, while
the hypothesis $m_d=2$ is already sufficiently good to be almost 
indistinguishable from further refinements (with the reasonable amount of
data points adopted in our simulations).

Nevertheless, as already anticipated, such an indicator does not necessarily
give a definite answer. In fact, we can imagine two distributions with
the same support but grossly different densities. A geometrical indicator
such as $\chi(m)$ would likely fail to identify at once such important 
differences since small distances would be found for all points (only a
finer analysis could possible allow detecting an insufficient quality of
the reconstruction). 

\begin{figure}
\psfig{file=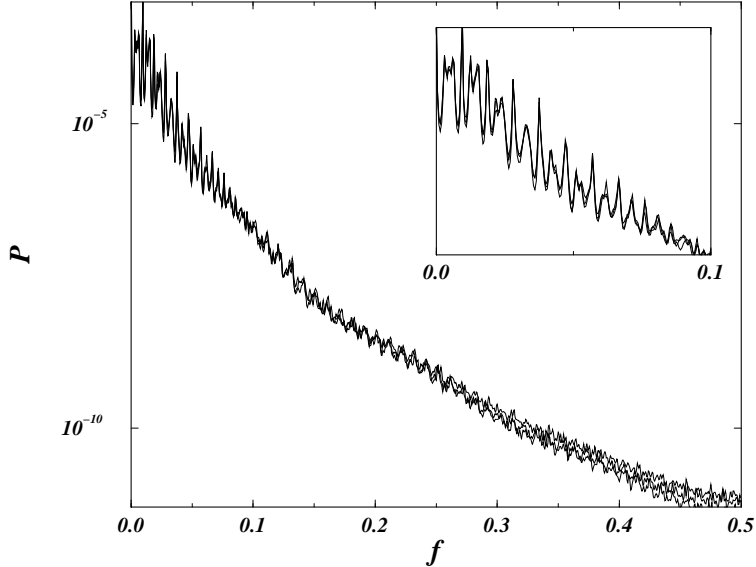,width=10cm,angle=270}
\caption{Power spectra of the original time series of the
generalized Mackey-Glass system,
the $(m=2)$-model, and the $(m=3)$-model.
The inset shows a blow-up for low frequencies.}
\label{fig.powerspec}
\end{figure} 

\begin{figure}
\psfig{file=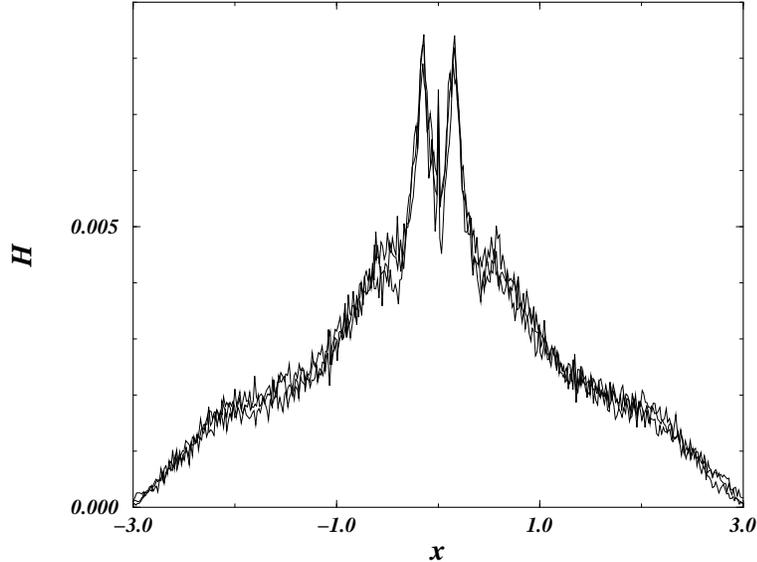,width=10cm,angle=270}
\caption{Histograms of the original time series of the
generalized Mackey-Glass system, 
the $(m=2)$-model, and the $(m=3)$-model.}
\label{fig.histogram}
\end{figure}

Therefore, we have decided to compute other quantities which have also
a direct physical meaning. In Fig.~\ref{fig.powerspec} and 
Fig.~\ref{fig.histogram}, we compare the spectra and the  histograms of 
the original time series and of the iterated time series $\hat{y}_2$,
$\hat{y}_3$. A good agreement is achieved in both cases. Since no 
significant improvements are found in going from $m=2$ to $m=3$, we can 
confirm the previous conjecture that $m_d=2$ is the minimal number of 
components necessary for a good reconstruction.

As a consequence of a successful modelling, it is not only possible to 
forecast the evolution in the real space, but also to extract information 
about the tangent space. In particular, one
can compute the Lyapunov spectrum (LS) \cite{Hegger99} for different 
choices of $m$ (in correspondence of the optimal value of the delay). 
We expect that the LS grossly differs from the correct one whenever $m$ 
is chosen too small, so that we can use Lyapunov exponents as a further 
global indicator to judge the quality of the reconstructed model. 

\begin{figure}
\psfig{file=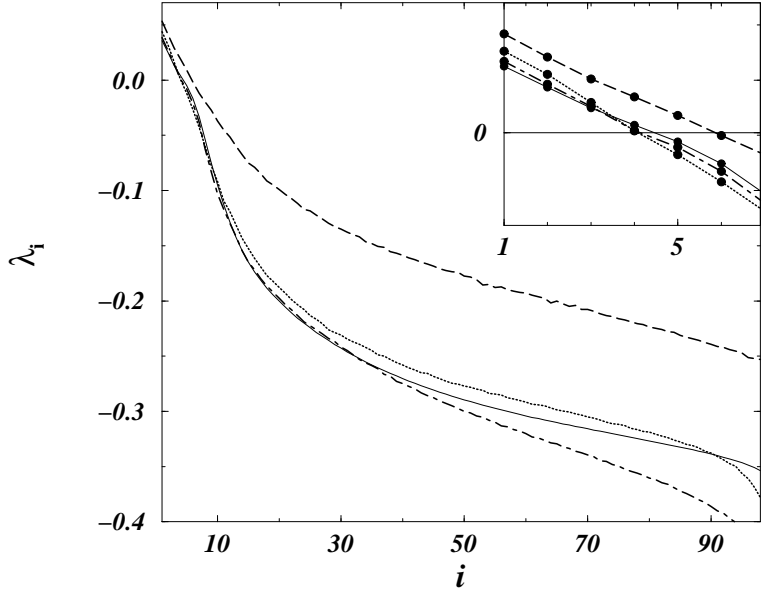,width=10cm,angle=270}
\caption{Lyapunov spectra of the generalized Mackey-Glass
system as estimated from the equations (solid line); 
from the $(m=1)$-model (dashed line); 
from the $(m=2)$-model (dotted line); 
from the $(m=3)$-model (dot-dashed line).
The inset shows a blow-up of the largest Lyapunov exponents.}
\label{fig.lyapspec}
\end{figure}

In our example of the generalized Mackey-Glass equation, we estimated the
LS for $m=1,2,$ and 3. The results are compared with the estimation of 
the spectrum obtained by direct integration of the equations
(see Fig. (\ref{fig.lyapspec})). Again we observe large deviations for $m=1$, 
while for $m \ge 2$, the LS is rather close to the true spectrum, thus
confirming once more the scenario suggested by the other indicators.

While it is not the goal of this paper to derive general quantitative
estimates of the various sources of errors, we would finally like to draw
the attention of the reader on the effect of dynamical noise that,
to different extent, affects any experiment. Dynamical noise, even
more strongly than measurement noise, can mask the presence of
hidden degrees of freedom. Once more, the Mackey-Glass system is a 
simple system to illustrate this phenomenon, since additional noise 
in the scalar model can induce jumps from positive to negative
values of the variable $y$, thus making the evolution essentially
indistinguishable from that of a noisy generalized system. 
As a consequence, the problem of a correct identification of the
deterministic components depends on the possible/required accuracy
of the modellization. In some cases, it might even be desirable
to model only the gross nonlinear features in terms of a few variables
while assimilating all the others to a sort of background noise 
indistinguishable from the true noise.

We conclude with a remark about the length of the time series.
It has been emphasized by some authors that the number of points 
necessary to estimate the delay time in a scalar system 
series can be quite small (say 500-1000) compared to the 
number of data points required in conventional nonlinear time
series analysis. In principle, we can confirm this result
for multi-component systems and a single time-delay feedback.
On the one hand, the discovery of hidden variables requires 
embedding the data in spaces of increasingly higher dimensions 
(though much smaller than the dimension of the entire phase 
space) and thus an increasingly larger number of data points.
On the other hand, since we do not aim at detecting scale 
invariant properties, the number of points required to obtain 
statistically significant results for the estimation of the 
delay time can be comparably small.

%% file: open.tex
In this paper we have shown that time-delayed feedback systems can be
investigated on the basis of a single valued time series. In particular, we
have seen that the dynamics can be reconstructed in low-dimensional
state-spaces even when the attractor dimension can be arbitrarily large.
This result has been obtained by restricting ourselves to the case where 
only one variable is fed back with a fixed specific delay time $\tau_0$. 
Two possible generalizations of this setup can be conceived that might be 
of interest in practical applications.

First, one can assume that the single feedback variable acts with several
different delays. As long as the number of such interactions is finite,
no dramtic changes are expected from a theoretical point of view:
instead of working with a ``two--window'' embedding, it should suffice to
to use an $(n+1)$--window embedding, where $n$ is the number of delays. This
is a straighforward generalization, as long as the windows do not overlap. 
Of course, the advantage of a low dimensionality of the state space is 
lost as soon as $n$ becomes large, but for only a few delays it might still 
work reasonably well. Completely different is the situation when we have to 
deal with a continuous spectrum of delays. In this case we expect this
method to fail, as it is no longer possible to reconstruct the equations of
motion in a low-dimensional manifold. 

A second possible generalization consists in sticking to a single delay 
time $\tau_0$, but admitting that several variables are fed back. This 
is similar to the case where we measure the {\it wrong} variable, as only a 
scalar variable is used to reconstruct the dynamics. This suggests that the 
length of the two windows should be increased by some factor in this case.

Another open problem concerns the uncertainty affecting a DM model that 
arises from the model mismatch due to the supposedly continuous-time dynamics.
In this paper, we have employed perturbative arguments to estimate the order of
magnitude of the FCE, when the windows are not too long. However, this
is still insufficient to draw mathematically rigorous conclusions about
the convergence properties of DM models towards the expected continuous-time
limit. In fact, a non-perturbative approach is presumably necessary to
deal with large window-lengths, besides the inclusion of additional 
information about the dynamical behaviour of the process under investigation.
This is a hard task that extends a general and still unsolved problem:
that of estimating the indeterminacy of an optimal prediction (on the basis
of the standard embedding approach) for a high-dimensional deterministic 
process.

%% file: acknow.tex
M. J. B. is supported by
a Marie-Curie-Fellowship of the EU with the contract number:
ERBFMBICT972305. R. H. is partly supported by the EU with the contract
number: ERBFMRXCT96.0010 and wants to thank the colleagues at the INO
for their kind hospitality.

%% file: mainfile.bbl
\begin{thebibliography}{10}

\bibitem{Packard80}
N.~H. Packard, J.~P. Crutchfield, J.~D. Farmer, and R.~S. Shaw, Phys. Rev.
  Lett. {\bf 45},  712  (1980).

\bibitem{Takens80}
F. Takens,  in {\em Dynamical Systems and Turbulence (Warwick 1980)}, Vol.~898
  of {\em Lecture Notes in Mathematics}, edited by D.~A. Rand and L.-S. Young
  (Springer-Verlag, Berlin, 1980), pp.\ 366--381.

\bibitem{Grassberger83d}
P. Grassberger and I. Procaccia, Phys. Rev. Lett. {\bf 50},  346  (1983).

\bibitem{Kantz97c}
H. Kantz and T. Schreiber, {\em Nonlinear Time Series Analysis} (Cambridge
  Univ. Press, Cambridge, UK, 1997).

\bibitem{Mackey77}
M.~C. Mackey and L. Glass, Science {\bf 197},  287  (1977).

\bibitem{Hale77}
J.~K. Hale, {\em Theory of functional differential equations} (Springer-Verlag,
  New York, 1977).

\bibitem{Bellmann63}
R. Bellmann and K.~L. Cooke, {\em Differential-Difference Equations} (Academic
  Press, New York, 1963).

\bibitem{Arecchi86}
F. Arecchi, W. Gadomski, and R. Meucci, Phys.\ Rev.\ A {\bf 34},  1617  (1986).

\bibitem{Farmer82}
J.~D. Farmer, Physica D {\bf 4},  366  (1982).

\bibitem{GiaLePo95}
G. Giacomelli, S. Lepri, and A. Politi, Physical Review E {\bf 51},  3131
  (1995).

\bibitem{Olbrich97}
E. Olbrich and H. Kantz, Phys. Lett. A {\bf 232},  63  (1997).

\bibitem{Fowler93}
A.~C. Fowler and G. Kember, Phys. Lett. A {\bf 125},  402  (1993).

\bibitem{Tian97}
Y. Tian and F. Gao, Physica D {\bf 108},  113  (1997).

\bibitem{Buenner96}
M.~J. B\"unner {\it et~al.}, Phys. Lett. A {\bf 211},  345  (1996).

\bibitem{BuennerPRE96a}
M.~J. B\"unner, T. Meyer, A. Kittel, and J. Parisi, Phys. Rev. E {\bf 54},
  R3082  (1996).

\bibitem{Voss97}
H. Voss and J. Kurths, Phys. Lett. A {\bf 234},  336  (1997).

\bibitem{Ellner97}
S. Ellner {\it et~al.}, Physica D {\bf 110},  182  (1997).

\bibitem{Hegger99}
R. Hegger, Estimation of the Lyapunov spectrum of time delay feedback systems,
  to appear in PRE, 1999.

\bibitem{BuennerEPL98}
M.~J. B\"unner {\it et~al.}, Europhys. Lett. {\bf 42},  353  (1998).

\bibitem{Hegger98}
R. Hegger, M. B\"unner, H. Kantz, and A. Giaquinta, Phys.\ Rev.\ Lett. {\bf
  211},  345  (1998).

\bibitem{Sauer91a}
T. Sauer, J.~A. Yorke, and M. Casdagli, J. Stat. Phys. {\bf 65},  579  (1991).

\bibitem{Casdagli92}
M. Casdagli,  in {\em Nonlinear Modeling and Forecasting, SFI Studies in the
  Sciences of Complexity} (Addison-Wesley, Reading, 1992).

\bibitem{SBDH97}
J. Stark, D. Broomhead, M. Davies, and J. Huke, Nonlinear Analysis, Methods \&
  Applications {\bf 30},  5303  (1997).

\bibitem{BuennerPRE97b}
M.~J. B\"unner, T. Meyer, A. Kittel, and J. Parisi, Phys. Rev. E. {\bf 56},
  5083  (1997).

\bibitem{Ikeda87}
K. Ikeda and K. Matsumoto, Physica D {\bf 29},  223  (1987).

\bibitem{He93a}
X. He and A. Lapedes, Physica D {\bf 70},  289  (1993).

\bibitem{Dorizzi87}
B. Dorizzi {\it et~al.}, Physical Review A {\bf 35},  328  (1987).

\bibitem{BaPo97}
R. Badii and A. Politi, {\em Complexity, hierarchical structures and scaling in
  physics}, {\em Cambridge Nonlinear Science Series} (Cambridge University
  Press, Cambridge, 1997).

\bibitem{Kantz94a}
H. Kantz, Phys. Rev. E {\bf 49},  5091  (1994).

\bibitem{Schreiber97}
T. Schreiber and A. Schmitz, Phys.\ Rev.\ Lett. {\bf 79},  1475  (1997).

\end{thebibliography}
